\documentclass[conference]{IEEEtran}
\IEEEoverridecommandlockouts
\usepackage{cite}
\usepackage{amsmath,amssymb,amsfonts}
\usepackage{algorithmic}
\usepackage{graphicx}
\usepackage{textcomp}
\usepackage{xcolor}
\usepackage{url}
\def\BibTeX{{\rm B\kern-.05em{\sc i\kern-.025em b}\kern-.08em
    T\kern-.1667em\lower.7ex\hbox{E}\kern-.125emX}}
\begin{document}

\title{From Framework to Practice: Designing a Real-World Telehealth Application for Palliative Care
\thanks{Email (corresponding author): rashina.hoda@monash.edu}}

\author{
\IEEEauthorblockN{Wei Zhou, Rashina Hoda\textsuperscript{*}, Andy Li, Chris Bain}
\IEEEauthorblockA{\textit{Faculty of Information Technology} \\
\textit{Monash University}\\
Clayton, VIC 3800, Australia}
\and
\IEEEauthorblockN{Laura Bird, Emmy Trinh, Peter Poon}
\IEEEauthorblockA{\textit{Faculty of Medicine, Nursing and Health Sciences / Monash Health} \\
\textit{Monash University / Monash Health} \\
Clayton, VIC 3800 / Clayton, VIC 3168, Australia}
\and
\IEEEauthorblockN{Teresa O'Brien, Mahima Kalla, Olivia Metcalf, Wendy Chapman}
\IEEEauthorblockA{\textit{Centre for Digital Transformation of Health} \\
\textit{University of Melbourne} \\
Parkville, VIC 3052, Australia}
\and
\IEEEauthorblockN{Joycelyn Ling}
\IEEEauthorblockA{\textit{Digital Health Cooperative Research Centre (DHCRC)} \\
Sydney, NSW 2000, Australia}
\and
\IEEEauthorblockN{Sam Georgy}
\IEEEauthorblockA{\textit{Healthdirect Australia} \\
Haymarket, NSW 2000, Australia}
\and
\IEEEauthorblockN{David Bevan}
\IEEEauthorblockA{\textit{Department of Health, Victoria} \\
Melbourne, VIC 3000, Australia}}

\maketitle

\begin{abstract}

As digital health solutions continue to reshape healthcare delivery, telehealth software applications have become vital for improving accessibility, continuity of care, and patient outcomes. This paper presents an analysis of designing a software application focused on Enhanced Telehealth Capabilities (ETHC) for palliative care, integrating across three socio-technical dimensions: quality, human values, and real-world. Designing for quality attributes -- such as performance, maintainability, safety, and security -- ensured that the system is technically robust and compliant with clinical standards. Designing for human values -- empathy, inclusivity, accessibility, and transparency -- helped enhance patient experience, trust, and ethical alignment. Designing for real-world -- through a multidisciplinary, experience-based co-design approach involving clinicians, patients, and carers that guided iterative cycles of prototyping, usability testing, and real-world evaluation -- ensured continuous refinement of features and alignment with clinical practice. The resulting telehealth software solution demonstrated that our socio-technical design framework was successful in producing a secure, equitable, and resilient digital health application. Our design approach can assist others designing software in health and other domains.

\end{abstract}






\maketitle

\section{Introduction}

Digital health technologies are reshaping healthcare delivery by enhancing accessibility, continuity of care, and patient outcomes \cite{dorsey2016state}. Telehealth software systems, in particular, have become indispensable tools that enable remote consultations, patient monitoring, and secure data sharing through scalable and flexible digital platforms \cite{gajarawala2021telehealth}. The COVID-19 pandemic accelerated this transformation, driving large-scale adoption and revealing the importance of designing sustainable, resilient telehealth infrastructures \cite{pertl2023recruitment, omboni2022worldwide}.

Despite these advances, designing telehealth software applications that are operationally sustainable, socially accessible, and clinically meaningful remains a complex challenge. This complexity arises from the need to balance requirements across multiple \textit{socio-technical} design dimensions -- \textit{quality}, \textit{human values}, and \textit{real-world}, ``\textit{where the social and technical aspects are interwoven}'' \cite[Chapter 3]{hoda2024qualitative}

\textbf{Designing for quality} concerns ensuring that telehealth applications meet the technical and professional standards required for safe and effective care. As outlined in the Software Engineering Body of Knowledge (SWEBOK) \cite{bourque2004software}, quality attributes such as performance, maintainability, safety, and security form the foundation for any healthcare software system. Existing research underscores that failure to embed these attributes during design can lead to performance bottlenecks, data integrity issues, and patient safety risks \cite{flott2021digital, volk2015safety, keller2017patient}. In digital health, quality also encompasses compliance with data protection and ethical governance frameworks to ensure transparency and accountability \cite{mcinerney2022patient, al2023ensuring}. Studies on privacy-preserving architectures, encryption mechanisms, and secure authentication highlight how robust implementation mitigates user mistrust and facilitates long-term system adoption. Therefore, designing for quality not only ensures system reliability but also builds the technical credibility necessary for clinical integration and regulatory acceptance.

\textbf{Designing for human values} extends beyond functional performance to address the human, ethical, and emotional aspects \cite{hussain2020human}. Digital health technologies are used by diverse stakeholders -- clinicians, patients, carers -- each with unique expectations, cognitive abilities, and cultural contexts. Research in human-computer interaction and participatory design emphasizes that embedding human values into technology leads to tools that are more trusted, inclusive, and effective \cite{o2017framework, marent2018ambivalence, wang2024designing, lupton2017digital}. Co-design and user-centered design methodologies are particularly powerful in ensuring that technologies reflect lived experience, rather than imposing one-size-fits-all solutions \cite{nusir2024systematic, tanprasert2024chi}. Empathy-driven design has been shown to enhance user satisfaction and adoption, especially in sensitive domains such as palliative and chronic care, where communication, dignity, and relational trust are paramount \cite{hussain2020human, gunatilake2024enablers}. Integrating accessibility and inclusivity into the design process ensures that applications do not merely serve clinical efficiency but also uphold fairness, accessibility, and respect for patients’ autonomy and emotional wellbeing.

\textbf{Designing for the real-world dimension} addresses the translation of research prototypes into sustainable, deployable, and maintainable systems that can function in diverse healthcare environments. While laboratory-based evaluations provide controlled insights, the complexity of real-world clinical workflows introduces challenges that require iterative, adaptive methods. Implementation research underscores that factors such as interoperability with existing electronic medical records, variation in clinical routines, and the digital literacy of users significantly affect success \cite{murray2016evaluating, bolinger2024key}. Frameworks combining co-design, simulation-based evaluation, and phased deployment can bridge this gap, ensuring that usability and technical performance align with actual clinical practice \cite{torous2019creating}. Moreover, sustained engagement with clinicians and patients after deployment -- through feedback loops and analytics -- supports ongoing adaptation, helping systems remain relevant amid evolving care models. This real-world perspective situates telehealth not as a standalone technological intervention but as an evolving part of a broader healthcare ecosystem.

Taken together, these three design dimensions -- \textit{quality}, \textit{human values}, and \textit{real-world} -- form \textit{an integrated socio-technical framework for designing digital health software} that balances engineering quality, ethical sensitivity, and contextual adaptability. We applied the framework to the design of the \textit{Patient Consultation Summary add-on}\footnote{\url{https://help.vcc.healthdirect.org.au/apps-tools/patient-consult-summary-application}}, developed in the Enhanced Telehealth Capabilities (ETHC) project. The add-on enables clinicians to create structured summaries of telehealth consultations in real-time, improving communication, patient comprehension, and care continuity. 

\begin{figure}[t]
    \centering
    \includegraphics[scale=0.17]{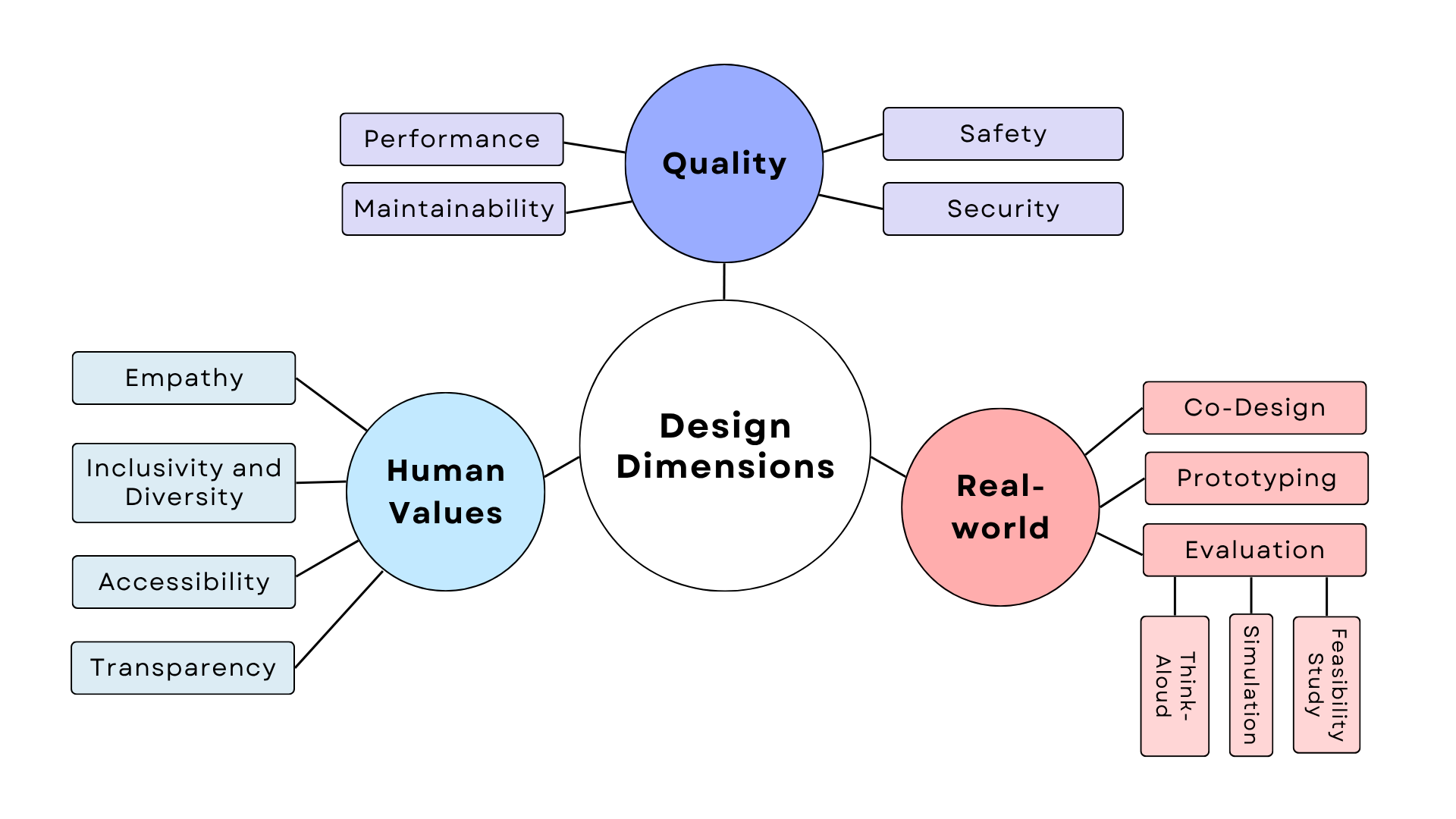}
    \caption{An integrated socio-technical framework for designing digital health software solutions}\label{fig1}
\end{figure}

By articulating these interrelated socio-technical dimensions, this paper aims to contribute a design-oriented perspective to digital health research, highlighting how multidisciplinary collaboration and iterative co-design can lead to solutions that are both technically robust and human-centered. The framework discussed here is not limited to telehealth but may be extended to broader digital health and societal applications that require the careful balancing of quality, human value attributes, and real-world feasibility. Fig. \ref{fig1} presents the integrated framework for digital health software design.

\section{Case study Context}\label{background}

The digital application designed in this project -- termed the Patient Consultation Summary add-on -- extends the practice of providing post-visit summaries, long used in physical clinical settings, into the telehealth domain, where such functionality has often been limited or non-existent \cite{corser2017after, lehnbom2014electronic}. 
The summary format can be set up to capture key details such as diagnosis, care plans, medication instructions, and follow-up requirements. The add-on is also designed to include simplified explanations of medical terminology sourced from a credible glossary with the option of clinicians adding their own definitions of terms during individual consultations.
By allowing the summary to be created, edited, and shared digitally and synchronously during the consultation, this solution enhances communication and comprehension supportive of better quality patient-centred care. 

This section outlines the key stakeholders involved, the clinical context in which the system was designed, and the technical integration that enabled its real-world deployment within a national telehealth platform.

\subsection{Project Stakeholders}

The project brought together a consortium of academic, clinical, and industry partners:  
\textit{Monash University} led the overall project, its coordination, and the software development;  
\textit{Monash Health} served as the clinical partner for pilot testing and real-world evaluation;  
\textit{Healthdirect Australia\footnote{\url{https://www.healthdirect.gov.au}} (HDA)} provided the national telehealth platform and technical integration environment;  
\textit{The University of Melbourne} contributed to co-design and simulation through the Digital Health Validitron\footnote{\url{https://www.melbconnect.com.au/community/the-digital-health-validitron}};  
and the \textit{Victorian Department of Health} offered policy and governance oversight.  
The\textit{ Digital Health Cooperative Research Centre (DHCRC)} supported collaboration, reporting, and knowledge translation between partners.

Stakeholder engagement was embedded throughout the project’s lifecycle -- from conceptualisation to feasibility evaluation.  
The multidisciplinary team included software engineering researchers, user-experience designers, clinicians, and healthcare researchers.  
Through participatory workshops, interviews, and iterative prototyping, clinicians, patients, and carers co-shaped the design.  
This participatory approach ensured the resulting solution aligned with real clinical workflows, privacy standards, and patient needs.

\subsection{Design Context: Palliative Care}

Palliative care, as defined by the World Health Organization, is a multidisciplinary approach aimed at improving the quality of life for patients and families facing life-threatening illnesses through the prevention and relief of suffering \cite{sepulveda2002palliative}. It involves early identification, assessment, and treatment of pain and other challenges - physical, psychosocial, and spiritual. This domain was chosen for its complexity and the need for comprehensive, holistic care, and best practices communication, making it a strong candidate for telehealth intervention with principles which may be applied across many healthcare settings. 
Patients requiring palliative care often experience mobility challenges, fluctuating symptoms, and high care needs \cite{de2022palliative}. Telehealth provides a means to reduce burdens on patients (e.g., unnecessary travel), while also enabling clinicians to maintain consistent support and monitoring.

\subsection{Platform Integration and Deployment}

To ensure scalability, sustainability, and trust, the application was integrated into the HDA platform, a national telehealth platform funded by the Australian government. It provides secure, encrypted video consultation capabilities via WebRTC (Web-based real time communitcation protocol) and supports multi-party video calls, including interpreters and carers into clinician-patient calls. 
Up to September 2025 end, the HDA Video Call service has facilitated over 7.1 million video call consults since the pandemic outbreak in March 2020. The HDA Video Call platform is used by over 120,100 health service providers in 12,700 health organisational clinics in over 90 specialties. 
The top clinical specialties (by consultation volumes) using the system include Emergency Medicine, Mental Health, General Practice, General Paediatrics, Oncology, and Neurology. The platform’s broad reach and existing clinical integration allowed for real-world evaluations and potential scaling.

\section{Designing for Quality
}\label{Quality}
In developing our telehealth application, we paid close attention to quality requirements to ensure the system not only meets functional demands but also performs efficiently, securely, and safely in clinical settings. According to the SWEBOK Guide, addressing quality requirements such as performance, maintainability, safety, and security is essential for building a robust and user-centric digital health solution. Below, we detail how our design addressed these critical areas.

\subsection{Performance}
To enhance the user experience and maintain the seamless operation of telehealth consultations, we incorporated strategies that ensured optimal system performance. Specifically:

\textit{Pre-loading Files:} We designed the system to pre-load all potential files required for a consultation session, including all the required file for using all the add-on.
Although this approach might appear redundant, it ensures a seamless consultation experience by eliminating interruptions caused by real-time file transfers, thereby minimizing bandwidth usage and reducing the risk of performance delays.

\textit{Backward Compatibility:} The code was developed with backward compatibility in mind, ensuring that updates do not disrupt existing functionalities. For versions where backward compatibility could not be maintained, we implemented backup solutions to preserve user data and functionality continuity.

\subsection{Maintainability}
Ensuring the long-term maintainability of the application was a key focus during the design phase. We implemented the following strategies:

\textit{File Management:} We structured the codebase to avoid oversized files while ensuring files were logically segmented by functionality. Each file was designed to support a standalone feature or render a specific page, with related files grouped into folders.

\textit{Code Management:} The overall code architecture was modularized to promote independent updates. This modular approach allows for isolated modifications without cascading impacts on other parts of the system, provided input/output structures remain consistent. Functions intended for cross-file usage were consolidated into a single module to prevent complex dependency chains.

\textit{Element ID Structure:} To facilitate debugging and enhance maintainability, we adopted a structured approach for element IDs. Instead of relying on auto-generated IDs, we manually created IDs using a meaningful format (e.g., “Prefix-Page-Section-UniqueID-suffix”) for easier tracking and debugging.

\subsection{Safety}
Safety was prioritized to ensure that the system operates without introducing clinical risks. Our approach to safety in the application encompassed:

\textit{Digital Safety:} The app was designed with a focus on technical robustness, minimizing risks such as data loss or unauthorized access. This ensures the system maintains integrity and security within a digital healthcare ecosystem.

\textit{Patient Safety:} The system was developed to support accurate information delivery and reliability. To mitigate risks related to human errors, we implemented a review process where doctors could verify the final version of patient summaries. This feature allows patients to review the information with their healthcare provider and ensures shared understanding.

\subsection{Security}
Security was an integral part of the design, given the sensitive nature of patient data. We focused on:

\textit{Data Privacy:} The system was designed with a serverless approach (in line with the HDA platform architecture) to safeguard patient data by processing information locally and avoiding the use of third-party servers.

\textit{Data Transfer Security:} When data transmission to a server was necessary, we minimized the exposure of sensitive information by transmitting only generic data and performing processing on the client side. For example, the application was designed to receive a comprehensive dictionary from the server, allowing the client-side to process terms without transmitting sensitive patient data. 

\textit{Compliance with Guidelines:} Our application adhered to the privacy and security standards outlined in the HDA Application Specification and Guidelines. We ensured that no personal identifiable or health information was stored post-consultation within the Video Call platform.

\section{Designing for Human Values}\label{HumanValue}
In the development of telehealth applications, embedding human values  within the design is crucial to ensure that the technology serves users in meaningful and ethical ways. Our approach focused on key values such as empathy, inclusivity, accessibility, and transparency \cite{hussain2020human}, integrating these principles to create a patient-centered and equitable system.

\subsection{Empathy}
Empathy is essential for designers and developers to understand the user's context \cite{gunatilake2025manifestations}, particularly in healthcare settings where users may experience vulnerability. By prioritizing empathy, we ensured that features were crafted to address users’ pain points and enhance their interactions with the app. A significant part of this commitment was reflected in the telehealth project, where experience-based co-design was implemented. Clinicians, patients, and carers were actively engaged to identify critical touchpoints that could enhance the telehealth experience. For instance, empathy-driven design improvements were made for palliative care patients by reducing cognitive load (e.g., patients constantly being able to see the doctor on video call and their view not being blocked by the summary) and ensuring the app met the emotional and practical needs of these individuals (e.g., patients being able to view, comment on, and confirm summaries in real time). 
This approach helped maintain patient dignity and comfort during telehealth consultations.

\subsection{Inclusivity and Diversity}
To serve a diverse user base effectively, it is imperative that telehealth applications cater to users of different ages, genders, abilities, and cultural backgrounds. The co-design process included insights from culturally and linguistically diverse (CALD) patients and clinicians, resulting in features that accommodated non-native English speakers. One such feature is real-time medical terminology explanations, which facilitates understanding during consultations. Moreover, including interpreters in co-design sessions ensured that the app was accessible and beneficial for diverse patient groups. This inclusivity enhances the app's usability across populations, fostering equitable access and patient-centered care.

\subsection{Accessibility}
Designing health apps with  accessibility in mind ensures that they contribute positively to society and promote equity. Our project emphasized improving healthcare accessibility for underserved populations, including patients in rural or remote areas. The use of serverless technologies reinforced our commitment to protecting patient privacy and security, highlighting an ethical approach to handling sensitive data. By focusing on these aspects, we aimed to support equitable health outcomes and align with broader social justice goals.

\subsection{Transparency}
Transparency is vital in building trust between the app and its users. Communicating clearly about how data is collected, used, and protected is an integral part of responsible design. Our approach to transparency included providing real-time feedback to patients during telehealth consultations, showing care plan summaries and terminology explanations in a user-friendly format. This not only informed patients about their health journey but also fostered trust by demonstrating that the app operates with integrity and openness. These efforts contribute to clearer communication and improved patient engagement, supporting better health outcomes.

\section{Designing for Real World}
\label{realWorld}

Following the establishment of the collaborative framework, the project adopted an agile and human-centered approach to research and development (R\&D) involving iterative and incremental cycles of \textit{co-design}, \textit{prototyping}, and \textit{evaluation} in close collaboration with clinicians, patients, and carers, ensuring that the application was thoroughly adapted to real-world clinical settings.

\subsection{Co-Design}
We utilized an experience-based co-design approach, photo elicitation interviews \cite{kalla2025understanding}, to capture the perspectives of clinicians, patients, and carers. The co-design was conducted within the Supportive and Palliative Care Unit of Monash Health, covering both metropolitan and regional areas with existing virtual care infrastructure. 

Specifically, photo elicitation was used to explore the lived experiences of end-users, including clinicians and patients. Real clinicians and patients participated in the design process by taking photos in response to prompts about their telehealth environment, challenges, and possible improvements. This method revealed nuanced insights into user experiences, highlighting elements like home care comfort, technical barriers, and emotional connections with clinicians. Analysis combined photo content with interview transcripts, identifying key themes that were visually represented through storyboards. Thought experiments encouraged participants to reflect on both the current and potential future uses of telehealth, helping to uncover evolving needs that informed forward-looking design elements.

Our design integrated recommendations from the co-design process, emphasizing user experience, interface design for optimal usability, and user-system interaction to meet these complex needs.

\subsection{Prototyping}

Led by the priorities of the participants, software prototypes were designed and developed directly on the industry partner HDA's user acceptance testing (UAT) platform to ensure seamless integration with the telehealth platform. The iterative design and development of prototypes prioritized features identified as essential through co-design, enabling participating users to experience and provide feedback on the features as they were being implemented. The software development effort was guided by HDA's design and implementation guidelines to maintain the same level of privacy, safety, and security offered by the telehealth platform and all its software add-ons/features.

\subsection{Evaluation}
The prototype was evaluated through a series of progressively realistic settings -- beginning with \textbf{think-aloud} co-design sessions \cite{wallace2002assertions}, advancing through \textbf{simulated testing}, and culminating in a real-world \textbf{feasibility study}. This staged approach enabled iterative refinement of the prototype while ensuring that design decisions were grounded in both usability evidence and clinical practice.

\subsubsection{Think-Aloud Sessions}

Think-aloud sessions enabled participants to verbalize their thoughts while interacting with the software prototype. Both administrators and real clinicians, along with patients, engaged in these sessions to test the telehealth summary-generation tool. Participants engaged in individual think-aloud exercises and simulated mock consultations, articulating challenges and questions they encountered in real-time. Usability Testing (UT) and User Acceptance Testing (UAT) sessions provided detailed feedback, which informed iterative refinements in the prototype development.

The insights gathered from these sessions highlighted user expectations, usability issues, and potential barriers, guiding improvements and establishing benchmarks for further usability testing in telehealth software.

\subsubsection{Simulation}

Simulation-based testing was conducted within the Digital Health Validitron, 
which offers a unified service for supporting the
development, design, validation, and evaluation of digital health software innovations in a simulated and controlled environment.
Simulations involved creating patient personas for sensitive settings like palliative care, with real clinicians interacting with simulated patients in mock consultations. The use of Validitron and simulated patients was important given the ethics of involving people with limited life expectancy for this preliminary testing. 
This method allowed the team to gather usability and acceptability feedback safely. 

Results from these simulations indicated areas for improvement, such as optimizing interface layouts for both clinicians and patients, refining voice and text functionalities, and enhancing user interface intuitiveness.

\subsubsection{Feasibility Study}

Finally, a real-world feasibility study was undertaken within five outpatient services at Monash Health, focusing on palliative and oncology care. This phase assessed the tool’s practicality, integration into clinical workflows, and its perceived value for both patients and clinicians in routine care. A mixed-methods design supported the evaluation, combining three complementary data streams:
\begin{itemize}
    \item \textbf{Self-report instruments:} Structured surveys captured user-reported outcomes such as perceived usefulness, ease of use, satisfaction, communication enhancement, and adoption willingness.
    \item \textbf{Artefact analysis:} Consultation summaries generated by the add-on were reviewed for communicative effectiveness, linguistic clarity, and patient actionability, including the role of embedded terminology support.
    \item \textbf{System analytics:} Digital telemetry recorded real-time usage of core functions -- terminology lookup, voice input, section editing, and summary sharing -- revealing engagement patterns across service types and user roles.
\end{itemize}

Ethics approvals were obtained for each evaluation phase, ensuring compliance with clinical research standards and data privacy requirements. Feedback from all stages informed successive iterations of the prototype, resulting in enhanced usability, patient safety, and clinical integration. This stepwise progression -- from exploratory think-aloud sessions to real-world feasibility testing -- provided a comprehensive understanding of the application’s usability, adoption potential, and fitness for clinical practice.

\section{Discussion}

The designing of the ETHC application highlights several key considerations for creating effective telehealth softwares, particularly in sensitive domains like palliative care. This section reflects on the challenges encountered, the strategies employed to address them, and the broader implications for design and innovation.

\subsection{Key Challenges and Strategies}

One of the significant challenges was ensuring that the ETHC application aligned with both functional and quality requirements while remaining adaptable to diverse clinical settings. Balancing technical robustness with user-centered design demanded iterative refinements driven by extensive user feedback. Co-design sessions proved instrumental in identifying and addressing the cognitive, emotional, and cultural needs of patients and clinicians, ensuring that the application was both intuitive and practical.

Another challenge lay in transitioning the application from prototype to real-world deployment. Clinical simulations and phased testing were crucial for identifying usability issues and optimizing the system's interface, features, and performance. Incorporating advanced features like the Web Speech API for voice recognition further demonstrated the importance of integrating cutting-edge technology with user-friendly design.

\subsection{Implications for Telehealth Design}

The ETHC application’s features (instant consultation summary with medical terminology explanations), highlight the value of enhancing patient engagement and health literacy in telehealth settings. By empowering patients with accessible and clear health information, the application fosters a collaborative approach to care, where patients are actively engaged in their health management.

The use of secure data-handling protocols and compliance with healthcare standards underscores the importance of prioritizing privacy and security in telehealth applications. These measures build user trust and ensure that the software meet the rigorous demands of sensitive healthcare contexts.

\subsection{Future Directions}

The lessons learned from the ETHC project pave the way for future telehealth design innovations. Expanding the application’s capabilities to integrate seamlessly with electronic health records could further enhance care coordination and streamline clinical workflows. Exploring machine learning algorithms for personalized patient recommendations could significantly advance the utility of such solutions. Broader adoption of co-design methodologies in telehealth development can help address the diverse needs of underserved populations, including CALD communities and rural patients. By fostering inclusivity and adaptability, future telehealth solutions can better align with the global push for equitable and accessible healthcare.

\section{Conclusion}

This case study illustrates how a socio-technical design framework -- spanning quality, human values, and real-world considerations -- can effectively guide the development of clinically grounded digital health software. Through the ETHC project, we demonstrated how an multidisciplinary, values-led design process translated conceptual principles into a functioning telehealth solution deployed within a national platform. Rather than focusing solely on technical achievement, the project highlighted how continuous engagement with clinicians, patients, and carers informed critical design trade-offs between safety, usability, and empathy.

Beyond its outcome, the project offers a transferable model for bridging research prototypes and clinical practice. Embedding co-design and iterative evaluation within a real-world infrastructure enabled the system to evolve alongside stakeholder needs, policy constraints, and technological advances. The experience underscores that sustainable design requires not only technical robustness but also institutional collaboration and adaptability.

\section*{Acknowledgment}
This research is supported by the Digital Health CRC Limited (DHCRC), Monash University, Healthdirect Australia, Monash Health, University of Melbourne (Digital Health Validitron), and Victorian Department of Health. DHCRC is funded under the Australian Commonwealth's Cooperative Research Centres (CRC) Program.

\bibliographystyle{IEEEtran}
\bibliography{ref}

\end{document}